\documentstyle{article}
\newcommand{\eqn}[1]{(\ref{#1})}

\newcommand{\ft}[2]{{\textstyle\frac{#1}{#2}}}

\renewcommand{\u}[1]{{\bar{#1}}}
\newcommand{\ba}{\left(\begin{array}}
\newcommand{\ea}{\end{array}\right)}

\newcommand{\M}{{\cal M}}
\newcommand{\R}{{\cal R}}
\newcommand{\T}{{\cal T}}

\renewcommand{\L}{{\cal L}}

\begin{document}
\thispagestyle{empty}
\begin{flushright}
KUL-TF-98/60\\
{\tt hep-th/9812220}\\
\today\\
\end{flushright}
\vskip3cm
\begin{center}
{\large {\bf Near Horizon Supergravity Superspace$^\dagger$}}
\vskip 1 cm
{\bf  Piet Claus}\\
\vskip.5cm
{\small
Instituut voor theoretische fysica, Katholieke Universiteit Leuven,\\
Celestijnenlaan 200D, B-3001 Leuven, Belgium}
\end{center}
\vskip 1.5cm
\centerline{\bf Abstract}
We present a construction of the superspace of maximally supersymmetric
$adS_{p+2}\times S^{d-p-2}$ near-horizon geometry based entirely on
the supergravity constraints of which the bosonic space is a solution.
Besides the geometric superfields, i.e.  the vielbeine and the
spinconnection, we also derive the isometries of the superspace together
with the compensating tangent space transformations to all orders in $\theta$.
\vfill
{\vfill\leftline{}\vfill
\footnoterule
\noindent
{\footnotesize
$\phantom{a}^\dagger$ Based on a talk given at the {\sl 
Mid-term TMR meeting on Quantum Aspects of Gauge Theories, Supersymmetry and
Unification}, Corfu, September 20--26, 1998, to appear in the
proceedings.}}

\newpage
\setcounter{page}{1}

\section{Introduction and motivation}
Brane solutions to supergravity theories have been considered \cite{GT} as
interpolating solutions between two maximally supersymmetric vacua,
i.e.~flat space related to the brane at infinity and $adS_{p+2}
\times S^{d-p-2}$, together with non-trivial forms, which is the
near-horizon geometry of the brane solution.  Both solutions can be
understood as superspaces. The algebraic fact that the (super)isometries
of the $adS_{p+2} \times S^{d-p-2}$ (super)space form an extended
(super)conformal group in $p+1$ dimensions is the basic argument in favor
of the $adS$/(S)CFT conjecture of Maldacena \cite{malda}, which was
originally proposed in connection to branes.  The conjecture relates
superstring theory/supergravity on $adS_{p+2}\times S^{d-p-2}$ to a
superconformal field theory which lives on the $p+1$ dimensional boundary
of the $adS$ space.  The most elaborated example is certainly Type IIB
string theory on $adS_5\times S^5$ related to 4-dimensional $N=4$ super
Yang-Mills theory.

One way to realise the superconformal field theory in $p+1$ dimensions is
to consider a super $p$-brane probe in the near-horizon $adS_{p+2}\times
S^{d-p-2}$ background.  The actions of such probes consist of a (modified)
`Dirac'-type term (for $D$-$p$-branes this is the Born-Infeld action) and
a Wess-Zumino term, both defined in terms of the pull-backs to the world
volume of the geometric superfields of the background, i.e.~the
supervielbeine and super $p+2$-form.  By construction these actions are
invariant under the {\it rigid} superisometries of the background and under
{\it local} symmetries, i.e.~diffeomorphisms of the world volume and
$\kappa$-symmetry.  By fixing the local symmetries of super $p$-brane probe
actions in their {\it own} near-horizon backgrounds, the complete set of
isometries of the background are realized non-linearly on the world volume
of the brane.  The bosonic part, which leads to a conformal field theory on
the brane was carried out in \cite{CKKTVP}.

To construct the superconformal theory on the brane, however, one needs to
know the complete supergeometry of the background.  Therefore, we have to
construct the geometric superfields to {\it all orders} in anticommuting
superspace coordinates $\theta$, because these higher order $\theta$ terms
will constitute the interactions on the world volume, i.e.~in the
superconformal field theory.  The general form of the super
$p$-brane actions in terms of the geometric superfields have been constructed
but the proof of invariances of the action only made use of the constraints 
that these superfields satisfy. Although the complete geometric superfields 
were known for flat superspace they  have only recently been constructed to 
all orders in $\theta$ for the near-horizon superspace 
$adS_{p+2}\times S^{d-p-2}$ \cite{MT,KRR,Bernard}.  These
constructions were based on a supercoset $G/H$ construction where $G$ is
the relevant extended superconformal group and the stability group $H$
is the {\it product} group $SO(p+1,1)\times SO(d-p-2)$.  Also the
superisometries have been obtained in a closed form to all orders in
$\theta$ in \cite{CK} and therefore in principle we can obtain the full
superconformal symmetry on the brane by gauge fixing the local
diffeomorphisms and $\kappa$-symmetry.

The complete geometric data of the background can also be
constructed entirely in supergravity superspace, for a certain class of
backgrounds. This class is characterised by the fact that all covariant
geometric superfields become covariantly constant.  Essentially this
restricts us to the above mentioned vacua, which have been shown to be
exact in the framework of supergravity \cite{KRaj}.  As expected
the supergravity superspace construction, i.e.~the solution of the
supertorsion and curvature constraints to all orders in $\theta$ yields a
completely equivalent description of the supergeometry as the coset
construction, which has been shown in \cite{C} for the 11-dimensional
vacua.  The translation between the two approaches by
interpreting the supergravity constraints as Maurer-Cartan equations for
the supergroup $G$ in \cite{CK}.

The supergravity superspace description of the background could be required
in some cases.  E.g.~the supercoset $G/H$ construction of the $adS_{p+2}
\times S^{d-p-2}$ superspace yields the metric as a product metric.
Sometimes it could be nice to have the metric in Cartesian coordinates, in
which the metric is manifestly invariant with respect to the directions
along the brane and to those transverse to the brane.  Especially the
$R$-symmetry of the extended superconformal group is manifest in these
coordinates \cite{CKKTVP}.  Interpreting the supergravity constraints as
Maurer Cartan equations in these coordinates yield a soft supergroup $G$
with structure functions, which are covariantly constant.  Fortunately the
construction in \cite{CK} does not require the structure functions to be
constant but only covariantly constant.  

Here we will extend the developement in \cite{C} and construct
the geometric superfields and the isometries only relying on supergravity
constraints and not refering to any coset construction.

\section{Supergravity in Superspace}
We will derive the superspace with coordinates\footnote{The indices are
split into curved ones $\Lambda = \{\mu,\dot\alpha\}$, with $\mu$ bosonic
and $\dot\alpha$ fermionic.  The flat indices are denoted as
$\u\Lambda = \{a,\alpha\}$.},
\begin{equation}
Z^\Lambda = \{x^\mu, \theta^{\dot \alpha}\}\,.
\end{equation}
Supergravity in superspace is described in terms of geometric
superfields, i.e. the supervielbeine and the spinconnection
\begin{equation}
E^{\u \Lambda} = d Z^\Lambda E_\Lambda{}^{\u \Lambda}(Z)\,,\qquad
\Omega_{\u \Lambda}{}^{\u \Sigma} = d Z^\Lambda \Omega_{\Lambda;\u
\Lambda}{}^{\u \Sigma}\,.
\end{equation}
We consider Lorentz superspaces where the spinconnection is determined
entirely in terms of $\Omega^{ab}$ \cite{CF}.
\par
{}From these geometric superfields we can derive the torsion and the
curvature
\begin{equation}
\T^{\u\Lambda} = d E^{\u \Lambda} - E^{\u \Sigma} 
\Omega_{\u \Sigma}{}^{\u \Lambda}\,,\qquad
\R_{\u \Lambda}{}^{\u\Sigma} = d \Omega_{\u\Lambda}{}^{\u \Sigma} -
\Omega_{\u \Lambda}{}^{\u \Pi}\, \Omega_{\u\Pi}{}^{\u\Sigma}\,.
\end{equation}
These covariant superfields satisfy the Bianchi-identities
\begin{equation}
D \T^{\u \Lambda} = - E^{\u \Sigma} \R_{\u \Sigma}{}^{\u
\Lambda}\,,\qquad
D \R_{\u \Lambda}{}^{\u\Sigma} = 0\,.
\label{BianchiTR}
\end{equation}
\par
The local symmetries of the supergravity theory in superspace are given by
\begin{itemize}
\item Superdiffeomorphisms (with parameters $\Xi^\Lambda$)
\begin{equation}
\delta Z^\Lambda = - \Xi^\Lambda (Z)\,, \qquad \delta \Phi_n = {\cal
L}_{\Xi} \Phi_n\,,
\label{diffs}
\end{equation}
where ${\cal L}_{\Xi}$ is the super Lie-derivative along the super
vectorfield $\Xi \equiv \Xi^\Lambda \frac \partial{\partial \Xi^\Lambda}$
and $\Phi_n$ a generic $n$-form.
\item Tangent-space rotations (with parameter $L^{ab}$)
\begin{eqnarray}
&&\delta E^{a} = E^{b} L_{b}{}^{a} (Z)\,,\qquad
\delta E^\alpha = - (\frac 14 L^{ab} (Z) \Gamma_{ab}
E)^\alpha\,,\nonumber\\
&&\delta \Omega^{ab} = d L^{ab} - L^{a}{}_{c} \Omega^{cb} +
\Omega^{a}{}_{c} L^{cb}\,.\label{tangent}
\end{eqnarray}
\end{itemize}
The geometric superfields and parameters are polynomials in $\theta$ with
$x$-dependent coefficients.  To obtain the fieldcontent and symmetries of
the supergravity theory, it is clear that not all these fields can be
independent.  Therefore one imposes covariant constraints on the
supertorsion and curvature to reduce the fieldcontent, consistent with the
Bianchi-identities \eqn{BianchiTR}.  
For all relevant theories the constraints are known.  The 11-dimensional
constraints e.g.~have been derived in \cite{CF} and the 10-dimensional
ones can be found in \cite{HW}. They yield on-shell supergravity in both
cases.  Besides the vielbeine and spinconnection
there are also a number of forms present which we ignore for the moment.

\section{The supergeometry and isometries near the horizon}
In this section we will solve the supergravity constraints and the super
Killing equations to all orders in $\theta$.  We will restrict to
superspaces with {\sl vanishing gravitino}.  Then the constraints read in
general
\begin{eqnarray}
\T^a &=& \ft12 E^\beta E^\alpha \T_{\alpha\beta}{}^a\,,\nonumber\\
\T^\alpha &=& E^\beta E^a \T_{a\beta}{}^\alpha\,,\nonumber\\
\R^{ab} &=& \ft12 E^d E^c R_{cd}{}^{ab} + \ft12 E^\beta E^\alpha
\R_{\alpha\beta}{}^{ab}\,,
\label{torsions}
\end{eqnarray}
where $R_{ab}{}^{cd}$ is the curvature of the bosonic space.

To obtain a complete superspace description we have to solve these
equations.  In general this can be done order by order in $\theta$, which
seems very tedious.  However we can apply the following trick.  Consider
the transformation
\begin{equation}
Z^\Lambda = \{x^\mu,\theta^{\dot\alpha}\} \rightarrow Z_t^\Lambda =
\{x^\mu, t\theta^{\dot\alpha}\}\,.
\label{trescaling}
\end{equation}
We denote superfields as functions of rescaled $\theta$'s with a subscript
$t$.  Taking the derivative with respect to $t$ of $E_t$ and $\Omega_t$
leads to the coupled first-order equations in $t$,
\begin{eqnarray}
\partial_t E^{\u \Lambda}_t &=& d (\theta^\alpha E_{t;\alpha}{}^{\u
\Lambda}) - \theta^{\dot\alpha} E_{t;{\dot\alpha}}{}^{\u \Delta} E^{\u
\Sigma}_t {\cal T}_{t;\u \Sigma\u \Delta}{}^{\u \Lambda} + E_t^{\u \Sigma}
\theta^{\dot \alpha} \Omega_{t;{\dot\alpha}\,\u \Sigma}{}^{\u \Lambda} -
\theta^{\dot\alpha} E_{t;{\dot\alpha}}{}^{\u \Sigma} \Omega_{t;\u\Sigma}
{}^{\u\Lambda}\,,\nonumber\\
\partial_t \Omega_t^{ab} &=& d(\theta^{\dot \alpha}
\Omega_{t;{\dot\alpha}}{}^{ab}) - \theta^{\dot\alpha}
E_{t;{\dot\alpha}}{}^{\Sigma} E_t^{\Lambda} {\cal R}_{t;\u \Lambda
\u\Sigma}{}^{ab} + \Omega^{ac}_t
\theta^{\dot\alpha}\Omega_{t;{\dot\alpha}\,c}{}^{b} - \theta^{\dot\alpha}
\Omega_{t {\dot\alpha}}{}^{a}{}_{c} \Omega_t^{cb}\,.
\label{dert}
\end{eqnarray}
To solve these equations we make the assumptions (gauge choices)
\begin{equation}
\theta^{\dot\alpha} E_{t;\dot\alpha}{}^{a} = \theta^{\dot\alpha}
\Omega_{t;{\dot\alpha}}{}^{ab} = 0\,,\qquad \theta^{\dot\alpha}
E_{t;\dot\alpha}{}^{\dot\alpha} = \theta^{\dot\alpha} e_{\dot\alpha}{}^{
\alpha}(x) \equiv \Theta^{\alpha}\,.
\label{anszats}
\end{equation}
We get using \eqn{torsions},
\begin{eqnarray}
\partial_t E^\alpha_t &=& (d\Theta + \ft14 \Omega_t\cdot
\Gamma\Theta)^\alpha
- \Theta^\beta E^a \T_{a\beta}{}^\alpha\,,\qquad
\partial_t E^{a}_t = \Theta^{\beta} E_t^{\alpha} \T_{\alpha\beta}{}^a
\,,\nonumber\\
\partial_t \Omega^{ab}_t &=& \Theta^\beta E_t^\alpha
\R_{\alpha\beta}{}^{ab}\,.
\label{tsystem}
\end{eqnarray}
These equations can be solved order by order in $t$, by taking multiple
derivatives w.r.t. $t$ and considering the initial conditions
\begin{equation}
E^\alpha_{(t=0)} = 0\,,\qquad
E^a_{(t=0)} = e^a(x)\,,\qquad
\Omega^{ab}_{(t=0)} = \omega^{ab}(x)\,,
\end{equation}
where $e^a$ and $\omega^{ab}$ are the vierbein and spinconnection of the
bosonic background.  The explicit solution is writen in closed form as
\begin{eqnarray}
E_t^\alpha &=& \left(\frac{\sinh t{\cal M}}{{\cal
M}}D\Theta\right)^\alpha\,,\qquad
E_t^a = e^a - \Theta^\alpha \T_{\alpha\beta}{}^a \left(
\frac {\sinh^2 t{\cal M}/2}{{\cal M}^2} D\Theta\right)^\beta
\,,\nonumber\\
\Omega^{ab}_t &=& \omega^{ab} - \Theta^\alpha \R_{\alpha\beta}{}^{ab}
\left( \frac {\sinh^2 t{\cal M}/2}{{\cal M}^2} D\Theta\right)^\beta
\,,
\label{ssol}
\end{eqnarray}
where
\begin{equation}
D\Theta^\alpha = (d\Theta + \ft14 \omega\cdot\gamma\Theta)^\alpha
- \Theta^\beta E^a \T_{a\beta}{}^\alpha\,,\quad ({\cal M}^2)^\alpha{}_\beta =
-\ft 14 (\Gamma_{ab})^\alpha{}_\beta \Theta^\gamma
\Theta\R_{\delta\beta}{}^{ab} -\T_{a\gamma}{}^\alpha \Theta^\gamma
\Theta^\delta \T_{\delta\beta}{}^a\,.
\end{equation}
One can easily checks that \eqn{anszats} is satisfied.
\par
Having derived the geometric superfields $E^{\u \Lambda}$ and $\Omega^{ab}$, 
we are also interested in the transformations which leave these fields
invariant. A specific subset of the symmetries \eqn{diffs}--\eqn{tangent} with
rigid parameters will leave the above superfields invariant, this subset are
the Killing supervector $\Xi^\Lambda$ and the compensating
tangent-space rotation $L^{ab}$.  We first derive the higher order $\theta$
terms. The super Killing equations are given by (vanishing
transformations of the solution)
\begin{eqnarray}
0 &=& \L_{\Xi} E^{\u\Lambda} + E^{\u \Sigma}
L_{\u\Sigma}{}^{\u \Lambda}
= D \tilde \Xi^{\u \Lambda} - \tilde \Xi^{\u \Delta} E^{\u \Sigma} {\cal T}_{\u
\Sigma\u\Delta}{}^{\u \Lambda}  + E^{\u\Sigma}\tilde L_{\u\Sigma}{}^{\u
\Lambda}\,,\\
0 &=& {\cal L}_{\Xi} \Omega^{ab} +  d L^{ab}
- L^{a}{}_{c} \Omega^{cb} + \Omega^{a}{}_{c} L^{cb}
= D \tilde L^{ab} - \tilde \Xi^{\u \Sigma} E^{\u \Lambda} \R_{\u \Lambda\u
\Sigma}{}^{ab}\,,
\label{superkillings}
\end{eqnarray}
where
\begin{eqnarray}
\tilde L_{\u \Sigma}{}^{\u\Lambda} &=&
\Xi^{\Lambda} \Omega_{\Lambda;\u\Sigma}{}^{\u\Lambda} +
L_{\u\Sigma}{}^{\u\Lambda}\,,\qquad
\tilde \Xi^{\u \Lambda} = \Xi^\Lambda E_\Lambda{}^{\u
\Lambda}\,,\nonumber\\
D \tilde \Xi^{\u \Lambda} &=& d \tilde \Xi^{\u\Lambda} - \tilde \Xi^{\u
\Sigma} \Omega_{\u\Sigma}{}^{\u\Lambda}\,,\qquad
D \tilde L^{ab} = d \tilde L^{ab} - \tilde L^{a}{}_{c} \Omega^{cb} +
\Omega^{a}{}_{c} \tilde L^{cb}\,.\label{covkillings}
\end{eqnarray}
It turns out that we can solve the equations for $\tilde \Xi^{\u\Lambda}$
and  $\tilde L^{ab}$ to all orders in $\theta$ and therefore we derive the
Killing supervector $\Xi^\Lambda$ and compensating transformation $L^{ab}$.
Now we again rescale $\theta$'s in the parameters and derive them with
respect to $t$.  Using the constraints \eqn{torsions}, the Killing
equations \eqn{superkillings} and also \eqn{anszats} we obtain again a
coupled set of first order equations in $t$,
\begin{eqnarray}
\partial_t \tilde \Xi^\alpha{}_t &=& -\Theta^\beta \tilde \Xi_t^a
\T_{a\beta}{}^\alpha + \ft14 (\tilde L\cdot \Gamma
\Theta)^\alpha\,,\qquad
\partial_t \tilde \Xi^a_t = -\Theta^\beta \tilde \Xi_t^\alpha
\T_{\alpha\beta}{}^a\,,\nonumber\\
\partial \tilde L_t^{ab} &=& -\Theta^\beta \tilde \Xi^\alpha
\R_{\alpha\beta}{}^{ab}\,.
\label{killtsystem}
\end{eqnarray}
This set of equations can be solved in terms of the initial conditions
\begin{equation}
\tilde \Xi^a_{(t=0)} = \tilde \xi^a = \xi^\mu e_\mu{}^a\,,\qquad
\tilde \Xi^\alpha_{(t=0)} = \tilde \epsilon^{\alpha} = \epsilon^{\dot \alpha}
{\cal K}(x)_{\dot\alpha}{}^\alpha
\,,\qquad \tilde L^{ab}_{(t=0)} = \ell^{ab} + \xi^\mu \omega_\mu^{ab}\,,
\label{killinit}
\end{equation}
where $\xi^\mu$, $\tilde \epsilon^\alpha$ and $\ell^{ab}$ are the
bosonic killing vector and spinor and the compensating tangent-space
rotation. They have been reviewed in various coordinates in
\cite{CK}. The spinor $\epsilon^{\dot\alpha}$ is a constant spinor
related to the Killing spinor by the $x$-dependent matrix ${\cal K}$.  The
complete solution to the super Killing equations can then be written again
in closed form as
\begin{eqnarray}
\tilde \Xi^\alpha_t &=& (\cosh t\M \tilde \epsilon)^\alpha
+ \left( \frac {\sinh \M}\M {\cal B}\Theta\right)^\alpha\,,\nonumber\\
\tilde \Xi^a &=& \tilde \xi^a - \Theta^\alpha \T_{\alpha\beta}^a \left(
\frac {\sinh \M}\M \tilde\epsilon + 2 \frac {\sinh^2 \M}{\M^2} {\cal
B}\Theta\right)^\beta\,,\nonumber\\
\tilde L^{ab} &=& \tilde \ell^{ab} - \Theta^\alpha \R_{\alpha\beta}^{ab}
\left( \frac {\sinh \M}\M \tilde \epsilon + 2 \frac {\sinh^2 \M}{\M^2} {\cal
B}\Theta\right)^\beta\,, \nonumber\\
({\cal B} \Theta)^\alpha &=& - \Theta^\beta \tilde \xi^a \T_{a\beta}^\alpha
- \ft14 (\tilde \ell\cdot \Gamma \Theta)^\alpha\,.
\end{eqnarray}
This completes the derivation of the ``covariant'' Killing superfields to
all orders in $\theta$, which are obtained at $t=1$.  In a special gauge,
the {\sl Killing spinor gauge} \cite{KRR}, one can easily `invert' the
covariant Killing superfields to the Killing supervector and compensating
tangent space parameter acting on the coordinates and supervielbeine.  The
Killing spinor gauge consists of choosing the $e(x)_{\dot\alpha}{}^\alpha$
in \eqn{anszats} to be ${\cal K}(x)_{\dot\alpha}{}^\alpha$, defined in
\eqn{killinit} by the Killing spinor of the bosonic geometry, hence its
name.  In this gauge we have the following simplification
\begin{equation}
E_\mu{}^a = e_\mu{}^a(x)\,,\qquad E_\mu{}^\alpha = 0\,,\qquad
\Omega_\mu^{ab}=\omega^{ab}_\mu(x)\,,
\end{equation}
i.e.~the bosonic background is not affected by higher order $\theta$
corrections. We derive the following explicit expressions
$\Xi^{\dot\alpha}$ and $\Xi^\mu$, which give the variations of the
superspace coordinates
\begin{eqnarray}
-\delta \theta^{\dot\alpha} &=& \Xi^{\dot\alpha} = \left[{\cal K}^{-1}
{\cal M} \coth {\cal M}
\,\epsilon\right]^{\dot\alpha} + ({\cal K}^{-1} {\cal B} \Theta)^{\dot \alpha}\,,\nonumber\\
-\delta x^\mu &=& -\Xi^\mu = \xi^\mu - \Theta^\alpha \T_{\alpha\beta}^a
\left[\left(\frac {\tanh {\cal M}/2}{{\cal M}}\right)\epsilon\right]^\beta
e_a{}^\mu \,
\end{eqnarray}
and also the compensating tangent space rotation can also be derived.
To have a complete description of the near-horizon superspace, we also have
to define the non-trivial superforms of the supergravity theory.  The
superforms ${\cal F}$ are given in supergravity see e.g.~\cite{CF,HW} for
the 11-dimensional and 10-dimensional supergravity. They can be written in
covariant form i.e.
\begin{equation}
{\cal F} = E^{\Lambda_n}\dots E^{\Lambda_1} F_{\Lambda_1\dots\Lambda_n}\,.
\end{equation}
For the near-horizon solutions these covariant components
$F_{\Lambda_1\dots\Lambda_n}$ are covariantly constant and do not get
higher order $\theta$ corrections \cite{KRaj}.  Therefore since we know the
bosonic geometry and forms and we have derived $E^{\u \Lambda}$ to all
orders in $\theta$ we know the complete superforms.  In principle these
forms have to be guessed in the supercoset approach, but they are then
uniquely defined by the fact that they have to satisfy the appropriate
Bianchi identity. This concludes the derivation of the complete near-horizon superspace, forms
and the superisometries in supergravity. Although the supercoset methods
used in \cite{MT}--\cite{CK} give a very nice description, one can also use
supergravity superspace only.

\vskip2mm
\noindent{\bf Acknowledgements} It is a pleasure to thank R.~Kallosh,
D.~Sorokin and A.~Van~Proeyen for useful discussions.  I would
also like to thank the organisers of the mid-term TMR meeting in Corfu
where part of this work was presented.  Work suported by the European
Commission TMR programme ERBFMRX-CT96-0045.
% % ---- Bibliography ----
%

\end{document}